\documentclass[%
 amsmath,amssymb,
 aps,
 superscriptaddress
]{revtex4-1}
\usepackage[dvipdfmx]{graphicx}
\usepackage{bm}

\usepackage[top=20truemm,bottom=20truemm,left=20truemm,right=20truemm]{geometry}
\usepackage{array}
\usepackage{color}
\usepackage{setspace}

\begin{document}
\setlength\textfloatsep{11pt}
\preprint{APS/123-QED}

\title{Magnetization Process of the $n$-type Ferromagnetic Semiconductor (In,Fe)As:Be Studied by X-ray Magnetic Circular Dichroism}

\author{S. Sakamoto}
\affiliation{Department of Physics, The University of Tokyo, Bunkyo-ku, Tokyo 113-0033, Japan}
\author{L. D. Anh}%
\affiliation{Department of Electrical Engineering and Information Systems, The University of Tokyo, Bunkyo-ku, Tokyo 113-8656, Japan}
\author{P. N. Hai}%
\affiliation{Department of Electrical Engineering and Information Systems, The University of Tokyo, Bunkyo-ku, Tokyo 113-8656, Japan}
\author{G. Shibata}
\affiliation{Department of Physics, The University of Tokyo, Bunkyo-ku, Tokyo 113-0033, Japan}
\author{\\Y. Takeda}
\affiliation{Synchrotron Radiation Research Unit, Japan Atomic Energy Agency (JAEA), Sayo-gun, Hyogo 679-5148, Japan}
\author{M. Kobayashi}
\affiliation{Department of Applied Chemistry, The University of Tokyo, Bunkyo-ku, Tokyo 113-8656, Japan}
\author{Y. Takahashi}
\affiliation{Department of Physics, The University of Tokyo, Bunkyo-ku, Tokyo 113-0033, Japan}
\author{T. Koide}
\affiliation{Photon Factory, Institute of Materials Structure Science, High Energy Accelerator Research Organization (KEK), Tsukuba, Ibaraki 305-0801, Japan}
\author{M. Tanaka}
\affiliation{Department of Electrical Engineering and Information Systems, The University of Tokyo, Bunkyo-ku, Tokyo 113-8656, Japan}
\author{A. Fujimori}
\affiliation{Department of Physics, The University of Tokyo, Bunkyo-ku, Tokyo 113-0033, Japan}



\date{\today}

\begin{abstract}
In order to investigate the mechanism of ferromagnetic ordering in the new $n$-type magnetic semiconductor (In,Fe)As co-doped with Be, we have performed X-ray absorption spectroscopy and X-ray magnetic circular dichroism (XMCD) studies of ferromagnetic and paramagnetic samples.
The spectral line shapes suggest that the ferromagnetism is intrinsic originating from Fe atoms incorporated into the Zinc-blende-type InAs lattice. The magnetization curves of Fe measured by XMCD were well reproduced by the superposition of a Langevin function representing superparamagnetic (SPM) behavior of nano-scale ferromagnetic domains and a $T$-linear function representing Curie-Weiss paramagnetism even much above the Curie temperatures. The data at 20 K showed a deviation from the Langevin behavior, suggesting a gradual establishment of macroscopic ferromagnetism on lowering temperature.
The existence of nano-scale ferromagnetic domains indicated by the SPM behavior suggests spatial fluctuations of Fe concentration on the nano-scale.


\end{abstract}

\pacs{Valid PACS appear here}
\maketitle


\section{Introduction}
Since the discovery of ferromagnetism in (In,Mn)As \cite{Munekata:1989aa} and (Ga,Mn)As \cite{Ohno:1996aa}, the Mn-doped III-V ferromagnetic semiconductors (FMS) have been intensively studied, driven by the scientific interest of how ferromagnetism is realized by such a small amount of magnetic impurities \cite{Jungwirth:2006aa, Sato:2010aa} and by future technological applications utilizing both the charge and spin degrees of freedom \cite{Koshihara:1997aa, Ohno:2000aa, Dietl:2014aa}. Although their electronic structures and the mechanism of the ferromagnetism still remain highly controversial \cite{Dietl:2000aa, Dietl:2001aa, Nishitani:2010aa, Okabayashi:2001aa, Burch:2006aa, ohya2011nearly}, it is widely believed that carriers mediate ferromagnetic interaction between spatially separated magnetic ions. 
This kind of ferromagnetism is called carrier-induced ferromagnetism, and its microscopic elucidation and utilization have been the main subjects in the  field of semiconductor spintronics. In the case of Mn-doped III-V compounds, however, there is a limitation that only $p$-type materials are available because the $\rm{Mn}^{2+}$ ions substituting the group-III elements work not only as local magnetic moments but also as acceptors providing the system with hole carriers. 
 
Recently, Hai $et$ $al.$ \cite{Nam-Hai:2012aa, Nam-Hai:2012ac, Nam-Hai:2012ab, Duc-Anh:2014aa, Sasaki:2014aa} have synthesized a new $n$-type III-V FMS, namely, Fe- and Be-codoped InAs. Usually, the substitution of Fe$^{3+}$ ions for the In sites does not provide charge carriers, resulting in Curie-Weiss paramagnetism \cite{Haneda:2000aa}. However, Hai $et$ $al$. have shown that by co-doping (In,Fe)As with Be atoms, which enter the interstitial sites and act as double donors, electrons are introduced into the system and mediate ferromagnetic interaction between Fe atoms. Here, unlike the Mn-doped materials, (In,Fe)As:Be has the advantage that one can control the concentration of magnetic ions and that of charge carriers independently.

In order to reveal the nature of the ferromagnetism of this new FMS, it is necessary to characterize the electronic and magnetic structure on the microscopic level. For this purpose, we have performed X-ray absorption spectroscopy (XAS) and X-ray magnetic circular dichroism (XMCD) measurements. XMCD is defined as the difference between the absorption of right- and left-handed circularly polarized X-rays and is a powerful method to probe the element-specific electronic and magnetic properties of compounds by utilizing the absorption edges of constituent elements. Furthermore, one can exclude extrinsic contributions to magnetism such as magnetic contaminations and the diamagnetic response of the substrate in the case of thin films and that of the host material in the case of FMS. Here, such diamagnetic contributions prohibit the extraction of the intrinsic paramagnetic component \cite{kobayashi:2010aa}, which is important to characterize the complex magnetism of FMS. In addition, XMCD sum rules \cite{carra:1993aa,thole:1992aa} make it possible to obtain the orbital and spin magnetic moments separately. A recent XMCD study \cite{kobayashi:2014aa} have shown that the orbital moment of (In,Fe)As is significantly larger than that of Fe metal. Since Fe$^{3+}$ ($d^5$) does not have orbital magnetic moment, the observed finite orbital magnetic moment would represent the charge-transfer $d^{6}\underline{L}$ states of $\rm Fe^{3+}$, where $\underline{L}$ denotes a ligand hole.

In the previous study \cite{Nam-Hai:2012ac}, ferromagnetic domains of $\sim$10 $\rm{\mu}$m were observed at sufficiently low temperatures well below Curie temperature ($T_{\rm C}$) by magneto-optical imaging, explaining the hysteretic behavior of magnetization. 
At high temperatures much above $T_{\rm C}$, however, magnetization still strongly depends on magnetic field, which implies the existence of rather small ferromagnetic domains of $\sim$nm acting as a superparamagnet. Under such circumstances, it is important to know how the ferromagnetism evolves as a function of temperature and magnetic field on a more microscopic level, i.e., on the nano-meter scale. In this paper, we have performed a careful analysis of XMCD data in order to address the issue of the evolution of magnetism from high to low temperatures.

\section{experiment}
Three samples $\rm{In}_{0.95}\rm{Fe}_{0.05}$As:Be (sample 1), In$_{0.9}$Fe$_{0.1}$As:Be (sample 2)
and $\rm{In}_{0.95}\rm{Fe}_{0.05}$As (sample 3) without Be were synthesized using the low-temperature molecular beam epitaxy (LT-MBE) method as follows. The 20-30 nm-thick (In,Fe)As layers were grown on InAs(001) substrates at $240^{\circ}$C after growing 50 nm-thick InAs buffer layers at $500^{\circ}$C. In order to prevent surface oxidation, the samples were covered by sub-nm-thick amorphous As capping layers. The amount of Be was estimated to be $2.6\times10^{19} \rm\ {cm}^{-3}$. Samples 1 and 2 were found to be ferromagnetic, whose Curie temperatures were 15 K and 40 K, respectively, determined by the Arrott plot of visible magnetic circular dichroism (vis-MCD) intensities, while sample 3 was paramagnetic. 
Careful sample characterizations by transmission electron microscopy (TEM), energy dispersive X-ray spectroscopy (EDX), and three dimensional atom probe (3DAP) proved that there were no Fe related precipitates or secondary phases in the samples \cite{Nam-Hai:2012ac}. 
Experiments were done at the undulator beamline BL-16A2 of Photon Factory (PF), High Energy Accelerator Research Organization (KEK). Temperature was varied from 20 K to 260 K and magnetic field from 0 T to 5 T. The total electron-yield (TEY) mode was used. The direction of the incident X-rays and magnetic field was perpendicular to the sample surfaces, and XMCD spectra were obtained by changing the polarization of X-rays while the direction of magnetic field was fixed. Backgrounds of the spectra were assumed to be the summation of a linear function and two-step inverse tangent functions representing the Fe $L_{2,3}$ edge jumps \cite{Chen:1995aa}.

\section{results}
\begin{figure}[!]
\begin{center}
\includegraphics[width=8.4cm]{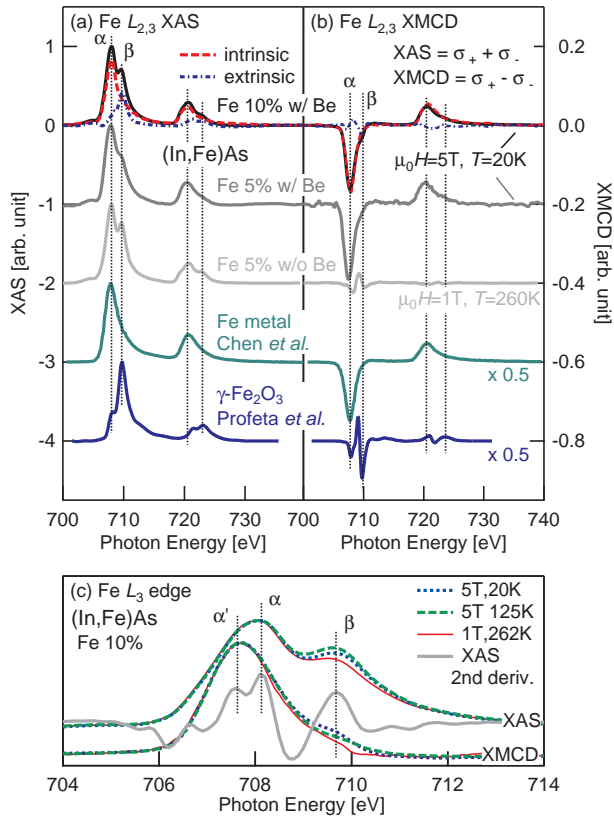}
\caption{XAS (a) and XMCD (b) spectra of (In,Fe)As in comparison with those of metallic Fe \cite{Chen:1995aa} and $\gamma$-$\rm{Fe_{2}O_{3}}$ \cite{BriceProfeta:2005aa}. The XAS and XMCD spectra have been normalized so that the peak intensities of XAS are equal to one. For the (In,Fe)As sample with 10\% Fe, the spectra have been decomposed into the intrinsic and extrinsic components as depicted with red dashed and blue dashed curves, respectively. (c) Fe $L_{3}$-edge XAS and XMCD spectra of the 10\% Fe-doped (In,Fe)As sample at different temperatures and magnetic fields. The XMCD spectra as well as XAS spectra have been normalized to its peak height. In order to emphasize weak structures, the second derivative of the XAS spectra with reversed sign are also shown by gray solid curve.}
\label{spectra}
\end{center}
\end{figure}

The XAS and XMCD spectra of (In,Fe)As with and without Be are shown in Fig. \ref{spectra} together with the spectra of Fe metal (ferromagnetic) and $\gamma$-Fe$_{\rm{2}}$O$_{\rm{3}}$ (ferrimagnetic) as references. Clear XMCD signals were obtained for the ferromagnetic (In,Fe)As samples and virtually no signal was detected for the paramagnetic (In,Fe)As sample as expected. The XAS spectra at the Fe $L_{3}$ edges are mainly composed of two structures $\alpha$ and $\beta$.
The small shoulder structure $\beta$ could be attributed to an extrinsic origin of Fe oxides at the surface \cite{kobayashi:2014aa}, considering that its energy is close to the peak energy of $\gamma$-Fe$_{2}$O$_{3}$ and that its intensity changes in non-systematic ways with varying magnetic field and temperature, as shown in Fig. \ref{spectra}(c). 
This probably resulted from the change of the beam position between different scans and inhomogeneous distribution of Fe oxides on the surface.
Utilizing this non-systematic spectral change, we have deduced intrinsic XAS and XMCD spectra in the following manner \cite{takeda:2008aa}.
First, using two spectra $\rm{S_{1}}$ and $\rm{S_{2}}$ with different intensities of structure $\beta$, the intrinsic spectrum $\rm{S_{int}}$ was obtained as
${\rm S_{int}} \propto {\rm S_{1}}- p{\rm S_{2}}$, where $p$ was chosen so that structure $\beta$ vanished. Then, the extrinsic spectrum $\rm{S_{ext}}$ was obtained as
${\rm S_{ext}} \propto {\rm S_{1}}- q {\rm S_{int}}$, where $q$ was chosen so that line shape of ${\rm S_{ext}}$ agreed with that of ${\gamma}$-$\rm{Fe_{2}O_{3}}$.
The decomposed intrinsic and extrinsic spectra for the 10\% Fe-doped sample are shown in Figs. \ref{spectra}(a) and \ref{spectra}(b). It was difficult to apply the same procedure to the 5\% Fe-doped sample because structure $\beta$ is less prominent and the intensity of structure $\beta$ did not change significantly between different scans. However, using the intrinsic and extrinsic spectra for the 10\% sample, we have estimated the mixing ratio of the intrinsic and extrinsic components, that is, how much oxides exist at the surface. Thus we could obtain the correct values of the magnetic moments for the 5\% Fe-doped sample as well as 10\% Fe-doped one using XMCD sum rules \cite{carra:1993aa,thole:1992aa}.
We note that, except for the extrinsic shoulder structures $\beta$, the line shapes of the XAS spectra were independent of the Fe and Be contents and shared a broad single-peak structure. This means that the local electronic structures of Fe atoms is basically the same between the three samples, notably between the ferromagnetic and paramagnetic ones. 
The result that multiplet structures reminiscent of the localized 3$d$ systems \cite{Laan:1992aa} are absent suggests that the $d$ electrons of Fe are relatively delocalized \cite{kang:2008aa, kowalik:2012aa} probably through the strong hybridization between the Fe 3$d$ orbitals and the ligand $s,p$ orbitals.

The deduced intrinsic XAS and XMCD spectra of (In,Fe)As:Be resemble those of Fe metal at a first glance, and therefore, one might suspect that the ferromagnetism originates from Fe metal precipitates. However, this possibility can be ruled out from following reasons. First, the XAS spectra of the paramagnetic sample showed the same spectral line shape as those of the ferromagnetic samples. If the observed spectral line shapes of (In,Fe)As were due to Fe metal particles, the XMCD intensity should have been comparable among the three (In,Fe)As samples including the paramagnetic one.
Second, looking at the magnified spectra of the Fe $L_{3}$ edge presented in Fig. \ref{spectra}(c), the peak positions of the XAS and XMCD do not coincide with each other unlike the XAS and XMCD spectra of Fe metal \cite{Saitoh:2012aa}. Corresponding to the peak in the XMCD spectra, there is a tiny shoulder $\alpha'$ in the XAS spectra $\sim$ 0.5 eV below its peak position $\alpha$. This shoulder is clearly seen in the second derivative of the XAS spectra with respect to photon energy, but less prominent in the spectra of Fe metal \cite{Saitoh:2012aa}. From these observations, we conclude that the ferromagnetism in this system is distinct from that of Fe metal and of the intrinsic origin arising from Fe atoms incorporated into the Zinc-blende-type InAs lattice.

In order to obtain the magnetic moment of Fe for the ferromagnetic samples, we have applied the XMCD sum rules \cite{carra:1993aa,thole:1992aa}:
\begin{eqnarray}
&\displaystyle m_{\rm orb}=-\frac{4\int_{L_{2,3}}\Delta\sigma\ {\rm d}\omega}{3\int_{L_{2,3}}\sigma\ {\rm d}\omega}n_{h},\\
&\displaystyle m_{\rm spin}+7m_{\rm T}=-\frac{6\int_{L_{3}}\Delta\sigma\ {\rm d}\omega-4\int_{L_{2,3}}\Delta\sigma\ \rm{d}\omega}{\int_{L_{2,3}}\sigma\ \rm{d}\omega}n_{h},\ \ \ \\
&\displaystyle\sigma=\sigma_{+}+\sigma_{-},\ \Delta\sigma=\sigma_{+}-\sigma_{-},
\end{eqnarray}
where $m_{\rm orb}$ and $m_{\rm spin}$ are the orbital and spin magnetic moments in units of $\mu_{\rm B}$/atom, respectively, and $m_{\rm T}$ is the expectation value of the magnetic dipole operator which is negligibly small for an atomic site with high symmetry such as $T_{d}$ or $O_{h}$ \cite{stohr:1995aa}.
$\sigma_{+}$ and $\sigma_{-}$ denote absorption cross sections for X-rays with positive and negative helicity, respectively, $n_{h}$ the number of 3$d$ holes.  We applied these equations to the decomposed intrinsic spectra instead of the raw spectra.
The area of the intrinsic XAS spectra, namely the denominator of the equations, were estimated to be $\sim$ 20\% smaller than that of the raw XAS spectra for the 10\% Fe-doped sample, and $\sim$ 5\% smaller for the 5\% Fe-doped one. 
As for the XMCD spectra, the extrinsic components were almost absent and did not contribute to the area of the XMCD spectra and hence to the numerator of Eqs. (1) and (2). 
Note that the existence of the extrinsic components in the XAS spectra only changes the calculated spin moment $m_{\rm spin}$ and orbital moment $m_{\rm orb}$ by the same factor while keeping the ratio $m_{\rm orb}/m_{\rm spin}$ unchanged.

\begin{table}[!]
\label{mom}
\begin{center}
\caption{Spin and orbital magnetic moments of Fe in (In,Fe)As:Be and Fe metal.}
\begin{tabular*}{8.5 cm}{l|ccc}
\hline\hline
\parbox[c][0.5cm][c]{0cm}& $m_{\rm orb}/m_{\rm spin}$ & $m_{\rm orb}$ & $m_{\rm spin}$ \\
\hline
\parbox[c][0.4cm][c]{0cm}{}$\rm{In_{0.95}Fe_{0.05}As}$:Be  & \ \ 0.088\ $\pm$\ 0.016\ \  & \ 0.22*\ \ & \ 2.49*\ \ \\
\parbox[c][0.4cm][c]{0cm}{}$\rm{In_{0.9}Fe_{0.1}As}$:Be  & \ \ 0.075 $\pm$ 0.015\ \  & \ 0.21*\ \  & \ 2.80*\ \  \\
\parbox[c][0.4cm][c]{0cm}{}$\rm{In_{0.95}Fe_{0.05}As}$:Be \cite{kobayashi:2014aa} & \ \ 0.065\ $\pm$\ 0.014\ \  & 0.10\ \  & \ 1.60\ \ \ \\
\parbox[c][0.4cm][c]{0cm}{}Fe bcc \cite{Chen:1995aa} & \ \ 0.043 $\pm$ 0.001\ \  & 0.085\ \  & \ 1.98\ \ \  \\
\hline\hline
\multicolumn{4}{r}{*measured at $\mu_{0}H$\rm = 5 T and $T$ = 20 K\ \ \ }
\end{tabular*}
\end{center}
\vspace{0cm}
\end{table}

\begin{figure}[]
\begin{center}
\includegraphics[width=8.4cm]{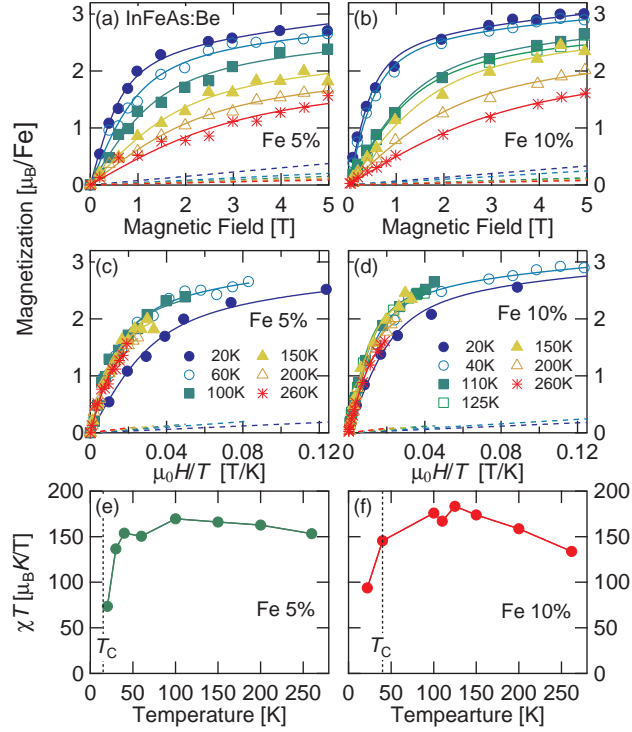}
\caption{XMCD intensities of (In,Fe)As:Be samples with 5\% and 10\% Fe doping. (a), (b) Magnetic field dependence of magnetization deduced from the Fe $L_{2,3}$-edge XMCD intensities at various temperatures. (c), (d) The same data as (a) and (b) plotted against $\mu_{0}H/T$. At the low temperature of 20 K, the magnetization shows a clear deviation from the high temperature data. In panel (a)-(d), solid curves represent the fitting curves by the summation of the superparamagnetic Langevin function and the paramagnetic linear function. The paramagnetic components are separately shown by dashed lines. (e), (f) Slope of the magnetization curves at zero magnetic field against $\mu_{0}H/T$, namely $\varDelta M/\varDelta\left(\frac{\mu_{0}H}{T}\right)|_{H\to 0}=\chi T$ as a function of temperature.}
\label{MH}
\end{center}
\end{figure}

Table I summarizes the orbital and spin moments of (In,Fe)As:Be deduced using the XMCD sum rules and those of bcc Fe metal \cite{Chen:1995aa}.
For the calculation of the moments, we assumed the number of electrons in the Fe 3$d$ states to be 5 and the correction factor for the spin sum rule to be 0.685 for Fe$^{3+}$ \cite{Piamonteze:2009aa}. Being consistent with the previous report \cite{kobayashi:2014aa}, the ratio of the orbital moment to the spin moment $m_{\rm orb}/m_{\rm spin}$, which is not affected by the assumption of $n_{h}=5$ and the existence of the extrinsic component in the XAS spectra, is larger than that of Fe metal and is positive. This can be explained if the $d^6\underline{L}$ configuration is mixed into the ground state due to charge transfer from ligand orbitals to the Fe 3$d$ orbitals.

\begin{figure}[!t]
\begin{center}
\includegraphics[width=5.5cm]{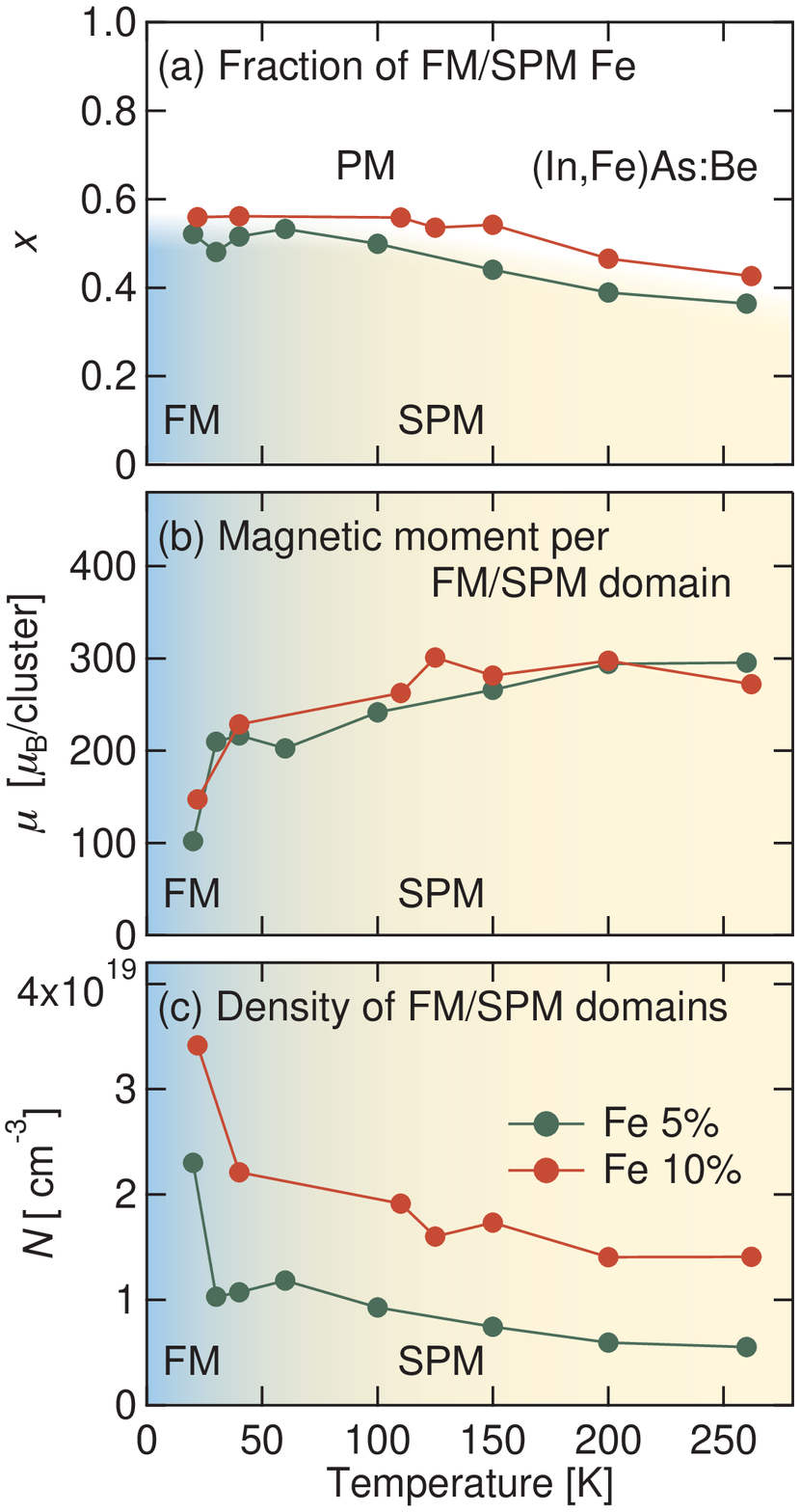}
\caption{(a) Fraction $x$ of Fe atoms participating in ferromagnetism or superparamagnetism to all the Fe atoms. (b) Total magnetic moment $\mu$ inside one FM/SPM domain, obtained by fitting Eq. (4) to the data. (c) Density of FM/SPM domain $N$ estimated from the fitting parameters $x$ and $\mu$.}
\label{parameters}
\end{center}
\end{figure}

Magnetization ($M$) per Fe atom at various magnetic fields and temperatures has thus been deduced from the XMCD intensities and are plotted in Figs. \ref{MH}(a) and \ref{MH}(b). Here, the horizontal axis is the effective magnetic field corrected for the demagnetizing field of the thin films although this correction is small in the case of the diluted magnetic systems. 
In the present experimental set-up, hysteresis, which was observed in the previous study \cite{kobayashi:2014aa}, was too small to be detected. It is worth noting that the $M$-$H$ curves were almost identical between 5\% and 10\% Fe-doped samples even though the $T_{\rm{C}}$'s were different.
At low magnetic fields, a steep increase of magnetization was observed even above $T_{\rm C}$ and, at high magnetic fields, linear increase of magnetization was observed even at the low temperature of 20 K where the ferromagnetic component would be saturated. 
This behavior can be understood if two contributions to the magnetization are superimposed, namely, superparamagnetism from nano-scale ferromagnetic domains (FM/SPM domains) and paramagnetism from isolated magnetic moments. Therefore, we have fitted the following summation of the Langevin function L($\xi$) and a $T$-linear function to the experimental XMCD intensity:
\begin{eqnarray}
&\displaystyle M=5x{\rm{L}}\left( \frac{\mu \mu_{0}H}{k_{\rm{B}}  T}\right)+\left(1-x\right)\frac{C\mu_{0}H}{T+T_{A}},\ \ \\
&\displaystyle \rm{L}\left( \xi\right)=\coth\left(\xi\right)-\frac{1}{\xi},
\end{eqnarray}
where $x$ $\left(0\leq x \leq 1\right)$ denotes the fraction of Fe atoms participating in the superparamagnetism or ferromagnetism, $\mu$ the total magnetic moment per nano-scale FM/SPM domain, $C$ the Curie constant of the $\rm{Fe^{3+}}$ ion with 5 $\mu_{\rm{B}}$, and $T_{A}$ the Weiss temperature representing the antiferromagenetic interaction between paramagnetic $\rm{Fe^{3+}}$ ions. 

Figures \ref{MH}(c) and \ref{MH}(d) show the magnetization curves at various temperatures replotted as functions of $\mu_{0}H/T$. They approximately fall onto a single curve, which is a fingerprint of superparamagnetism, except for the data at 20 K close to $T_{\rm C}$. The deviation may represent the onset of macroscopic ferromagnetism at 20 K. Figures \ref{MH}(e) and \ref{MH}(f) show the slopes of magnetization at zero magnetic field with respect to $\mu_{0}H/T$, i.e. $\varDelta M/\varDelta\left(\frac{\mu_{0}H}{T}\right)|_{H\to 0} =\chi T$. If the system follows the Langevin law of superparamagneitsm, $\chi T$ should be constant. Hence the drop of $\chi T$ around 50 K on decreasing temperature indicates the gradual establishment of macroscopic ferromagnetism from the nano-scale superparamagnetism.

The $M$-$H$ data were fitted by Eq. (4) assuming $T_{\rm A} = 30$ K, which is nearly equal to the Weiss temperature of paramagnetic (Ga,Fe)As without carrier doping (= 32 K) \cite{Haneda:2000aa}, and the results of the fit are shown in Figs. \ref{MH}(a)-\ref{MH}(d) by solid curves. In the figures, the paramagnetic components are separately shown by dashed lines. Note that a finite value of $T_{\rm A}$ larger than 30 K was necessary to fit to the data.  
Figures \ref{parameters}(a) and \ref{parameters}(b) show the temperature dependence of fitting parameters $x$ and $\mu$. 
The fraction of Fe atoms participating in the ferromagnetism or superparamagnetism ($x$) is estimated to be 40\%-60\% and increasing as the temperature is decreased. 
The magnetic moment per FM/SPM domain ($\mu$) are 200-300 $\mu_{\rm{B}}$, which corresponds to 40-60 Fe atoms, and they decrease as the temperature is decreased. 
Interestingly, the estimated values are very similar between the two samples. 
The estimated density of FM/SPM domains $N$ are shown in Fig. \ref{parameters}(c) and found to be $\sim10^{19}$ cm$^{-3}$, comparable to the density of Be atoms.

\section{discussion}

\begin{figure}[!t]
\begin{center}
\includegraphics[width=8.4cm]{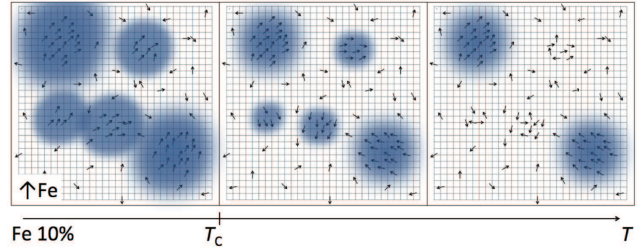}
\caption{Schematic picture of the formation of nano-scale FM domains in (In,Fe)As:Be. There exist FM Fe-rich domains and paramagnetic isolated Fe ions. Here the density of Fe atoms and the size and the number of FM domains reflect the obtained value for the 10\% Fe-doped sample.}
\label{scm}
\end{center}
\end{figure}

There may be two possible scenarios to explain the formation of nano-scale FM/SPM domains observed for (In,Fe)As. 
One is the bound magnetic polaron (BMP) model \cite{kaminski:2002aa, coey:2005aa}, where relatively localized carriers stabilize ferromagnetic spin alignment extending over the scale of nm.
In this model, the polarons overlap with each other and form macroscopic ferromagnetic domains at the critical concentration where the system satisfies the Mott criterion for insulator-to-metal transition.
Second is the formation of Fe rich regions during the growth while maintaining the zinc-blende crystal structure, often discussed in the context of spinodal decomposition \cite{sato:2005aa, kuroda:2007aa}, as schematically shown in Fig. \ref{scm}.
Considering the metallic behavior of (In,Fe)As \cite{Nam-Hai:2012ac}, in particular, the small effective mass of electrons in the conduction band \cite{Nam-Hai:2012ab}, the BMP model is less likely because it predicts a heavy effective mass for the charge carriers, and the second scenario is more probable.
As noted above, the $M$-$H$ behaviors are almost identical between the two ferromagnetic samples with different Fe contents, meaning one FM/SPM domain consists of 40-60 Fe atoms for both samples. 
If the first scenario is correct and the distribution of Fe atoms is rather uniform, electrons are bound to charged impurities such as ${\rm Be^{2+}}$ to form magnetic polarons and there should be difference in the total magnetic moment of a FM/SPM domain between the two samples because the number of Fe atoms within the same Bohr radius of electrons should be different. 
On the other hand, in the case of the second scenario, where Fe-rich FM/SPM domains are formed in the InAs host, the observed independence of magnetic moment on the Fe content can be explained if Fe-rich regions with optimized number of Fe atoms are formed. 
Suppose an extreme case where Fe atoms occupy all the In sites inside the FM/SPM regions, the radius of the domain containing 40-60 Fe atoms is estimated to be $\sim$ 1 nm, which is difficult to detect by conventional x-ray diffraction. 
From the fact that the density of FM/SPM domains are comparable to that of Be atoms, one can speculate that the FM Fe-rich domains were preferentially formed around Be impurities. Actually, it was reported that the co-doping of shallow impurities affects the distribution or aggregation of magnetic ions \cite{dietl:2014ab}.

We shall discuss below how the magnetism in this system behaves with temperature.
As can be seen from Figs. \ref{parameters}(a) and \ref{parameters}(b), the lower the temperature becomes, the more Fe atoms participate in ferromagnetism but the smaller the average size of FM/SPM domains becomes. This behavior can be understood if there exists size distribution of Fe rich regions. As described in Fig. \ref{scm}, at high temperatures, only large Fe-rich regions are ferromagnetic. On lowering temperature, smaller Fe-rich regions becomes ferromagnetic as described in the middle panel of Fig. \ref{scm}, and hence the average magnetic moment of FM/SPM domains ($\mu$) decreases [Fig. \ref{parameters}(b)]. On the other hand, the fraction of Fe atoms participating in ferromagnetism or superparamagnetism ($x$) increases [Fig. \ref{parameters}(a)] and the density of FM/SPM domains ($N$) increases [Fig. \ref{parameters}(c)]. 
Further decreasing the temperature below $T_{\rm C}$ would result in the overlap of the FM/SPM domains as shown in the left panel of Fig. \ref{scm}, and the macroscopic ferromagnetism, which cannot be described by the Langevin law, starts to be stabilized.
The reason for the drop of $\chi T$ at 20 K [Figs. \ref{MH}(e) and \ref{MH}(f)] instead of a divergence with the emergence of macroscopic ferromagnetism may be that the ferromagnetism is soft in (In,Fe)As. For example, let us write the magnetization of a system where superparamagnetism and soft ferromagnetism coexist as 
\begin{equation}
\frac{M(H)}{M_{sat}}=(1-y)L(\frac{\mu \mu_{0}H}{k_{\rm B}T})+y\tanh(H/H_{0}), 
\end{equation}
where the first term and second term represent the superparamagnetism and the soft ferromagnetism, respectively. $y$ denotes the fraction of ferromagnetism and $H_{0}$ is the saturation field. The second term is just an empirical expression \cite{coey:2010aa}, but one can assume that the magnetization is independent of temperature as long as the temperature is sufficiently low. Using this expression, $\chi T$ can be written as 
\begin{equation}
\frac{\chi T}{M_{sat}}= (1-y)\frac{\mu_{0}\mu}{3k_{\rm B}}+y\frac{T}{H_{0}}.
\vspace{1.0mm}
\end{equation}
Therefore, if $\mu_{0}\mu/3k_{\rm B}$ is larger than $T/H_{0}$, i.e. $H_{0}>3k_{\rm B}T/\mu_{0}\mu$, $\chi T$ would show a drop with the emergence of ferromagnetic term.

At the present stage, one can conclude that the large magnetization of the present (In,Fe)As:Be samples above $T_{\rm C}$ originates from the nano-scale FM domains. 
The previously observed ferromagnetic domain of $\sim$10 ${\rm \mu}$m at sufficiently low temperature would contain a large number of ($\sim 10^{7}$ assuming the domain size of $10{\rm \mu m}\times10{\rm \mu m}\times20 {\rm nm}$) the nano-scale FM domains revealed in the present work, and would result from ferromagnetic coupling between the FM/SPM domains and/or magnetic anisotropy of the FM/SPM domains.

\section{conclusion}
We have performed XMCD measurements on the new $n$-type ferromagnetic semiconductors (In,Fe)As:Be with different Fe and Be contents. Despite the fact that $T_{\rm{C}}$ was different between the samples, the XAS and XMCD line shapes and magnetization data were almost identical. The  analysis of the spectral line shapes and the magnetic moment using sum rules suggest that the ferromagnetism is intrinsic originating from the Fe atoms incorporated into the Zinc-blende-type InAs lattice, and not from segregated secondary phases. 
From the magnetization curves at various temperatures even much above $T_{\rm C}$, a SPM-like steep increase in the low $H$ region and a paramagnetic linear increase in the high $H$ region were observed, which implies the co-existence of FM domains of $\sim$nm size and paramagnetic isolated Fe$^{3+}$ ions. The fraction of Fe atoms participating in the ferromagnetism or superparamagnetism was estimated to be 40-60\% and the magnetic moment per FM/SPM domain to be 200-300 $\mu_{\rm B}$. 
The formation of the FM/SPM nano-domains can be explained by the formation of Fe-rich regions during the material growth possibly around Be atoms.

\section*{acknowledgments}
This work was supported by Grants-in-Aids for Scientific Research from the JSPS (No. S22224005 and 23000010).
The experiment was done under the approval of the Photon Factory Program Advisory Committee (proposal Nos. 2013S2-004 and 2012G667).
SS, LDA, GS, and YT acknowledge support from the Program for Leading Graduate Schools, LDA, GS and MK from the JSPS Fellowship for Young Scientists. PNH acknowledges support from the Yazaki Memorial Foundation for Science and Technology, the Murata Science Foundation, and the Toray Science Foundation.

\bibliography{bibtexFile}

\begin{thebibliography}{38}%
\makeatletter
\providecommand \@ifxundefined [1]{%
 \@ifx{#1\undefined}
}%
\providecommand \@ifnum [1]{%
 \ifnum #1\expandafter \@firstoftwo
 \else \expandafter \@secondoftwo
 \fi
}%
\providecommand \@ifx [1]{%
 \ifx #1\expandafter \@firstoftwo
 \else \expandafter \@secondoftwo
 \fi
}%
\providecommand \natexlab [1]{#1}%
\providecommand \enquote  [1]{``#1''}%
\providecommand \bibnamefont  [1]{#1}%
\providecommand \bibfnamefont [1]{#1}%
\providecommand \citenamefont [1]{#1}%
\providecommand \href@noop [0]{\@secondoftwo}%
\providecommand \href [0]{\begingroup \@sanitize@url \@href}%
\providecommand \@href[1]{\@@startlink{#1}\@@href}%
\providecommand \@@href[1]{\endgroup#1\@@endlink}%
\providecommand \@sanitize@url [0]{\catcode `\\12\catcode `\$12\catcode
  `\&12\catcode `\#12\catcode `\^12\catcode `\_12\catcode `\%12\relax}%
\providecommand \@@startlink[1]{}%
\providecommand \@@endlink[0]{}%
\providecommand \url  [0]{\begingroup\@sanitize@url \@url }%
\providecommand \@url [1]{\endgroup\@href {#1}{\urlprefix }}%
\providecommand \urlprefix  [0]{URL }%
\providecommand \Eprint [0]{\href }%
\providecommand \doibase [0]{http://dx.doi.org/}%
\providecommand \selectlanguage [0]{\@gobble}%
\providecommand \bibinfo  [0]{\@secondoftwo}%
\providecommand \bibfield  [0]{\@secondoftwo}%
\providecommand \translation [1]{[#1]}%
\providecommand \BibitemOpen [0]{}%
\providecommand \bibitemStop [0]{}%
\providecommand \bibitemNoStop [0]{.\EOS\space}%
\providecommand \EOS [0]{\spacefactor3000\relax}%
\providecommand \BibitemShut  [1]{\csname bibitem#1\endcsname}%
\let\auto@bib@innerbib\@empty
\bibitem [{\citenamefont {Munekata}\ \emph {et~al.}(1989)\citenamefont
  {Munekata}, \citenamefont {Ohno}, \citenamefont {von Molnar}, \citenamefont
  {Segm\"uller}, \citenamefont {Chang},\ and\ \citenamefont
  {Esaki}}]{Munekata:1989aa}%
  \BibitemOpen
  \bibfield  {author} {\bibinfo {author} {\bibfnamefont {H.}~\bibnamefont
  {Munekata}}, \bibinfo {author} {\bibfnamefont {H.}~\bibnamefont {Ohno}},
  \bibinfo {author} {\bibfnamefont {S.}~\bibnamefont {von Molnar}}, \bibinfo
  {author} {\bibfnamefont {A.}~\bibnamefont {Segm\"uller}}, \bibinfo {author}
  {\bibfnamefont {L.~L.}\ \bibnamefont {Chang}}, \ and\ \bibinfo {author}
  {\bibfnamefont {L.}~\bibnamefont {Esaki}},\ }\href {\doibase
  10.1103/PhysRevLett.63.1849} {\bibfield  {journal} {\bibinfo  {journal}
  {Phys. Rev. Lett.}\ }\textbf {\bibinfo {volume} {63}},\ \bibinfo {pages}
  {1849} (\bibinfo {year} {1989})}\BibitemShut {NoStop}%
\bibitem [{\citenamefont {Ohno}\ \emph {et~al.}(1996)\citenamefont {Ohno},
  \citenamefont {Shen}, \citenamefont {Matsukura}, \citenamefont {Oiwa},
  \citenamefont {Endo}, \citenamefont {Katsumoto},\ and\ \citenamefont
  {Iye}}]{Ohno:1996aa}%
  \BibitemOpen
  \bibfield  {author} {\bibinfo {author} {\bibfnamefont {H.}~\bibnamefont
  {Ohno}}, \bibinfo {author} {\bibfnamefont {A.}~\bibnamefont {Shen}}, \bibinfo
  {author} {\bibfnamefont {F.}~\bibnamefont {Matsukura}}, \bibinfo {author}
  {\bibfnamefont {A.}~\bibnamefont {Oiwa}}, \bibinfo {author} {\bibfnamefont
  {A.}~\bibnamefont {Endo}}, \bibinfo {author} {\bibfnamefont {S.}~\bibnamefont
  {Katsumoto}}, \ and\ \bibinfo {author} {\bibfnamefont {Y.}~\bibnamefont
  {Iye}},\ }\href {\doibase http://dx.doi.org/10.1063/1.118061} {\bibfield
  {journal} {\bibinfo  {journal} {Appl. Phys. Lett.}\ }\textbf {\bibinfo
  {volume} {69}},\ \bibinfo {pages} {363} (\bibinfo {year} {1996})}\BibitemShut
  {NoStop}%
\bibitem [{\citenamefont {Jungwirth}\ \emph {et~al.}(2006)\citenamefont
  {Jungwirth}, \citenamefont {Sinova}, \citenamefont {Ma\ifmmode~\check{s}\else
  \v{s}\fi{}ek}, \citenamefont {Ku\ifmmode~\check{c}\else \v{c}\fi{}era},\ and\
  \citenamefont {MacDonald}}]{Jungwirth:2006aa}%
  \BibitemOpen
  \bibfield  {author} {\bibinfo {author} {\bibfnamefont {T.}~\bibnamefont
  {Jungwirth}}, \bibinfo {author} {\bibfnamefont {J.}~\bibnamefont {Sinova}},
  \bibinfo {author} {\bibfnamefont {J.}~\bibnamefont {Ma\ifmmode~\check{s}\else
  \v{s}\fi{}ek}}, \bibinfo {author} {\bibfnamefont {J.}~\bibnamefont
  {Ku\ifmmode~\check{c}\else \v{c}\fi{}era}}, \ and\ \bibinfo {author}
  {\bibfnamefont {A.~H.}\ \bibnamefont {MacDonald}},\ }\href {\doibase
  10.1103/RevModPhys.78.809} {\bibfield  {journal} {\bibinfo  {journal} {Rev.
  Mod. Phys.}\ }\textbf {\bibinfo {volume} {78}},\ \bibinfo {pages} {809}
  (\bibinfo {year} {2006})}\BibitemShut {NoStop}%
\bibitem [{\citenamefont {Sato}\ \emph {et~al.}(2010)\citenamefont {Sato},
  \citenamefont {Bergqvist}, \citenamefont {Kudrnovsk\'y}, \citenamefont
  {Dederichs}, \citenamefont {Eriksson}, \citenamefont {Turek}, \citenamefont
  {Sanyal}, \citenamefont {Bouzerar}, \citenamefont {Katayama-Yoshida},
  \citenamefont {Dinh}, \citenamefont {Fukushima}, \citenamefont {Kizaki},\
  and\ \citenamefont {Zeller}}]{Sato:2010aa}%
  \BibitemOpen
  \bibfield  {author} {\bibinfo {author} {\bibfnamefont {K.}~\bibnamefont
  {Sato}}, \bibinfo {author} {\bibfnamefont {L.}~\bibnamefont {Bergqvist}},
  \bibinfo {author} {\bibfnamefont {J.}~\bibnamefont {Kudrnovsk\'y}}, \bibinfo
  {author} {\bibfnamefont {P.~H.}\ \bibnamefont {Dederichs}}, \bibinfo {author}
  {\bibfnamefont {O.}~\bibnamefont {Eriksson}}, \bibinfo {author}
  {\bibfnamefont {I.}~\bibnamefont {Turek}}, \bibinfo {author} {\bibfnamefont
  {B.}~\bibnamefont {Sanyal}}, \bibinfo {author} {\bibfnamefont
  {G.}~\bibnamefont {Bouzerar}}, \bibinfo {author} {\bibfnamefont
  {H.}~\bibnamefont {Katayama-Yoshida}}, \bibinfo {author} {\bibfnamefont
  {V.~A.}\ \bibnamefont {Dinh}}, \bibinfo {author} {\bibfnamefont
  {T.}~\bibnamefont {Fukushima}}, \bibinfo {author} {\bibfnamefont
  {H.}~\bibnamefont {Kizaki}}, \ and\ \bibinfo {author} {\bibfnamefont
  {R.}~\bibnamefont {Zeller}},\ }\href {\doibase 10.1103/RevModPhys.82.1633}
  {\bibfield  {journal} {\bibinfo  {journal} {Rev. Mod. Phys.}\ }\textbf
  {\bibinfo {volume} {82}},\ \bibinfo {pages} {1633} (\bibinfo {year}
  {2010})}\BibitemShut {NoStop}%
\bibitem [{\citenamefont {Koshihara}\ \emph {et~al.}(1997)\citenamefont
  {Koshihara}, \citenamefont {Oiwa}, \citenamefont {Hirasawa}, \citenamefont
  {Katsumoto}, \citenamefont {Iye}, \citenamefont {Urano}, \citenamefont
  {Takagi},\ and\ \citenamefont {Munekata}}]{Koshihara:1997aa}%
  \BibitemOpen
  \bibfield  {author} {\bibinfo {author} {\bibfnamefont {S.}~\bibnamefont
  {Koshihara}}, \bibinfo {author} {\bibfnamefont {A.}~\bibnamefont {Oiwa}},
  \bibinfo {author} {\bibfnamefont {M.}~\bibnamefont {Hirasawa}}, \bibinfo
  {author} {\bibfnamefont {S.}~\bibnamefont {Katsumoto}}, \bibinfo {author}
  {\bibfnamefont {Y.}~\bibnamefont {Iye}}, \bibinfo {author} {\bibfnamefont
  {C.}~\bibnamefont {Urano}}, \bibinfo {author} {\bibfnamefont
  {H.}~\bibnamefont {Takagi}}, \ and\ \bibinfo {author} {\bibfnamefont
  {H.}~\bibnamefont {Munekata}},\ }\href {\doibase 10.1103/PhysRevLett.78.4617}
  {\bibfield  {journal} {\bibinfo  {journal} {Phys. Rev. Lett.}\ }\textbf
  {\bibinfo {volume} {78}},\ \bibinfo {pages} {4617} (\bibinfo {year}
  {1997})}\BibitemShut {NoStop}%
\bibitem [{\citenamefont {Ohno}\ \emph {et~al.}(2000)\citenamefont {Ohno},
  \citenamefont {Chiba}, \citenamefont {Matsukura}, \citenamefont {Omiya},
  \citenamefont {Abe}, \citenamefont {Dietl}, \citenamefont {Ohno},\ and\
  \citenamefont {Ohtani}}]{Ohno:2000aa}%
  \BibitemOpen
  \bibfield  {author} {\bibinfo {author} {\bibfnamefont {H.}~\bibnamefont
  {Ohno}}, \bibinfo {author} {\bibfnamefont {D.}~\bibnamefont {Chiba}},
  \bibinfo {author} {\bibfnamefont {F.}~\bibnamefont {Matsukura}}, \bibinfo
  {author} {\bibfnamefont {T.}~\bibnamefont {Omiya}}, \bibinfo {author}
  {\bibfnamefont {E.}~\bibnamefont {Abe}}, \bibinfo {author} {\bibfnamefont
  {T.}~\bibnamefont {Dietl}}, \bibinfo {author} {\bibfnamefont
  {Y.}~\bibnamefont {Ohno}}, \ and\ \bibinfo {author} {\bibfnamefont
  {K.}~\bibnamefont {Ohtani}},\ }\href@noop {} {\bibfield  {journal} {\bibinfo
  {journal} {Nature}\ }\textbf {\bibinfo {volume} {408}},\ \bibinfo {pages}
  {944} (\bibinfo {year} {2000})}\BibitemShut {NoStop}%
\bibitem [{\citenamefont {Dietl}\ and\ \citenamefont
  {Ohno}(2014)}]{Dietl:2014aa}%
  \BibitemOpen
  \bibfield  {author} {\bibinfo {author} {\bibfnamefont {T.}~\bibnamefont
  {Dietl}}\ and\ \bibinfo {author} {\bibfnamefont {H.}~\bibnamefont {Ohno}},\
  }\href {\doibase 10.1103/RevModPhys.86.187} {\bibfield  {journal} {\bibinfo
  {journal} {Rev. Mod. Phys.}\ }\textbf {\bibinfo {volume} {86}},\ \bibinfo
  {pages} {187} (\bibinfo {year} {2014})}\BibitemShut {NoStop}%
\bibitem [{\citenamefont {Dietl}\ \emph {et~al.}(2000)\citenamefont {Dietl},
  \citenamefont {Ohno}, \citenamefont {Matsukura}, \citenamefont {Cibert},\
  and\ \citenamefont {Ferrand}}]{Dietl:2000aa}%
  \BibitemOpen
  \bibfield  {author} {\bibinfo {author} {\bibfnamefont {T.}~\bibnamefont
  {Dietl}}, \bibinfo {author} {\bibfnamefont {H.}~\bibnamefont {Ohno}},
  \bibinfo {author} {\bibfnamefont {F.}~\bibnamefont {Matsukura}}, \bibinfo
  {author} {\bibfnamefont {J.}~\bibnamefont {Cibert}}, \ and\ \bibinfo {author}
  {\bibfnamefont {D.}~\bibnamefont {Ferrand}},\ }\href {\doibase
  10.1126/science.287.5455.1019} {\bibfield  {journal} {\bibinfo  {journal}
  {Science}\ }\textbf {\bibinfo {volume} {287}},\ \bibinfo {pages} {1019}
  (\bibinfo {year} {2000})}\BibitemShut {NoStop}%
\bibitem [{\citenamefont {Dietl}\ \emph {et~al.}(2001)\citenamefont {Dietl},
  \citenamefont {Ohno},\ and\ \citenamefont {Matsukura}}]{Dietl:2001aa}%
  \BibitemOpen
  \bibfield  {author} {\bibinfo {author} {\bibfnamefont {T.}~\bibnamefont
  {Dietl}}, \bibinfo {author} {\bibfnamefont {H.}~\bibnamefont {Ohno}}, \ and\
  \bibinfo {author} {\bibfnamefont {F.}~\bibnamefont {Matsukura}},\ }\href
  {\doibase 10.1103/PhysRevB.63.195205} {\bibfield  {journal} {\bibinfo
  {journal} {Phys. Rev. B}\ }\textbf {\bibinfo {volume} {63}},\ \bibinfo
  {pages} {195205} (\bibinfo {year} {2001})}\BibitemShut {NoStop}%
\bibitem [{\citenamefont {Nishitani}\ \emph {et~al.}(2010)\citenamefont
  {Nishitani}, \citenamefont {Chiba}, \citenamefont {Endo}, \citenamefont
  {Sawicki}, \citenamefont {Matsukura}, \citenamefont {Dietl},\ and\
  \citenamefont {Ohno}}]{Nishitani:2010aa}%
  \BibitemOpen
  \bibfield  {author} {\bibinfo {author} {\bibfnamefont {Y.}~\bibnamefont
  {Nishitani}}, \bibinfo {author} {\bibfnamefont {D.}~\bibnamefont {Chiba}},
  \bibinfo {author} {\bibfnamefont {M.}~\bibnamefont {Endo}}, \bibinfo {author}
  {\bibfnamefont {M.}~\bibnamefont {Sawicki}}, \bibinfo {author} {\bibfnamefont
  {F.}~\bibnamefont {Matsukura}}, \bibinfo {author} {\bibfnamefont
  {T.}~\bibnamefont {Dietl}}, \ and\ \bibinfo {author} {\bibfnamefont
  {H.}~\bibnamefont {Ohno}},\ }\href {\doibase 10.1103/PhysRevB.81.045208}
  {\bibfield  {journal} {\bibinfo  {journal} {Phys. Rev. B}\ }\textbf {\bibinfo
  {volume} {81}},\ \bibinfo {pages} {045208} (\bibinfo {year}
  {2010})}\BibitemShut {NoStop}%
\bibitem [{\citenamefont {Okabayashi}\ \emph {et~al.}(2001)\citenamefont
  {Okabayashi}, \citenamefont {Kimura}, \citenamefont {Rader}, \citenamefont
  {Mizokawa}, \citenamefont {Fujimori}, \citenamefont {Hayashi},\ and\
  \citenamefont {Tanaka}}]{Okabayashi:2001aa}%
  \BibitemOpen
  \bibfield  {author} {\bibinfo {author} {\bibfnamefont {J.}~\bibnamefont
  {Okabayashi}}, \bibinfo {author} {\bibfnamefont {A.}~\bibnamefont {Kimura}},
  \bibinfo {author} {\bibfnamefont {O.}~\bibnamefont {Rader}}, \bibinfo
  {author} {\bibfnamefont {T.}~\bibnamefont {Mizokawa}}, \bibinfo {author}
  {\bibfnamefont {A.}~\bibnamefont {Fujimori}}, \bibinfo {author}
  {\bibfnamefont {T.}~\bibnamefont {Hayashi}}, \ and\ \bibinfo {author}
  {\bibfnamefont {M.}~\bibnamefont {Tanaka}},\ }\href {\doibase
  10.1103/PhysRevB.64.125304} {\bibfield  {journal} {\bibinfo  {journal} {Phys.
  Rev. B}\ }\textbf {\bibinfo {volume} {64}},\ \bibinfo {pages} {125304}
  (\bibinfo {year} {2001})}\BibitemShut {NoStop}%
\bibitem [{\citenamefont {Burch}\ \emph {et~al.}(2006)\citenamefont {Burch},
  \citenamefont {Shrekenhamer}, \citenamefont {Singley}, \citenamefont
  {Stephens}, \citenamefont {Sheu}, \citenamefont {Kawakami}, \citenamefont
  {Schiffer}, \citenamefont {Samarth}, \citenamefont {Awschalom},\ and\
  \citenamefont {Basov}}]{Burch:2006aa}%
  \BibitemOpen
  \bibfield  {author} {\bibinfo {author} {\bibfnamefont {K.~S.}\ \bibnamefont
  {Burch}}, \bibinfo {author} {\bibfnamefont {D.~B.}\ \bibnamefont
  {Shrekenhamer}}, \bibinfo {author} {\bibfnamefont {E.~J.}\ \bibnamefont
  {Singley}}, \bibinfo {author} {\bibfnamefont {J.}~\bibnamefont {Stephens}},
  \bibinfo {author} {\bibfnamefont {B.~L.}\ \bibnamefont {Sheu}}, \bibinfo
  {author} {\bibfnamefont {R.~K.}\ \bibnamefont {Kawakami}}, \bibinfo {author}
  {\bibfnamefont {P.}~\bibnamefont {Schiffer}}, \bibinfo {author}
  {\bibfnamefont {N.}~\bibnamefont {Samarth}}, \bibinfo {author} {\bibfnamefont
  {D.~D.}\ \bibnamefont {Awschalom}}, \ and\ \bibinfo {author} {\bibfnamefont
  {D.~N.}\ \bibnamefont {Basov}},\ }\href {\doibase
  10.1103/PhysRevLett.97.087208} {\bibfield  {journal} {\bibinfo  {journal}
  {Phys. Rev. Lett.}\ }\textbf {\bibinfo {volume} {97}},\ \bibinfo {pages}
  {087208} (\bibinfo {year} {2006})}\BibitemShut {NoStop}%
\bibitem [{\citenamefont {Ohya}\ \emph {et~al.}(2011)\citenamefont {Ohya},
  \citenamefont {Takata},\ and\ \citenamefont {Tanaka}}]{ohya2011nearly}%
  \BibitemOpen
  \bibfield  {author} {\bibinfo {author} {\bibfnamefont {S.}~\bibnamefont
  {Ohya}}, \bibinfo {author} {\bibfnamefont {K.}~\bibnamefont {Takata}}, \ and\
  \bibinfo {author} {\bibfnamefont {M.}~\bibnamefont {Tanaka}},\ }\href@noop {}
  {\bibfield  {journal} {\bibinfo  {journal} {Nat. Phys.}\ }\textbf {\bibinfo
  {volume} {7}},\ \bibinfo {pages} {342} (\bibinfo {year} {2011})}\BibitemShut
  {NoStop}%
\bibitem [{\citenamefont {Hai}\ \emph {et~al.}(2012{\natexlab{a}})\citenamefont
  {Hai}, \citenamefont {Sasaki}, \citenamefont {Anh},\ and\ \citenamefont
  {Tanaka}}]{Nam-Hai:2012aa}%
  \BibitemOpen
  \bibfield  {author} {\bibinfo {author} {\bibfnamefont {P.~N.}\ \bibnamefont
  {Hai}}, \bibinfo {author} {\bibfnamefont {D.}~\bibnamefont {Sasaki}},
  \bibinfo {author} {\bibfnamefont {L.~D.}\ \bibnamefont {Anh}}, \ and\
  \bibinfo {author} {\bibfnamefont {M.}~\bibnamefont {Tanaka}},\ }\href
  {\doibase http://dx.doi.org/10.1063/1.4730955} {\bibfield  {journal}
  {\bibinfo  {journal} {Appl. Phys. Lett.}\ }\textbf {\bibinfo {volume}
  {100}},\ \bibinfo {eid} {262409} (\bibinfo {year}
  {2012}{\natexlab{a}})}\BibitemShut {NoStop}%
\bibitem [{\citenamefont {Hai}\ \emph {et~al.}(2012{\natexlab{b}})\citenamefont
  {Hai}, \citenamefont {Anh}, \citenamefont {Mohan}, \citenamefont {Tamegai},
  \citenamefont {Kodzuka}, \citenamefont {Ohkubo}, \citenamefont {Hono},\ and\
  \citenamefont {Tanaka}}]{Nam-Hai:2012ac}%
  \BibitemOpen
  \bibfield  {author} {\bibinfo {author} {\bibfnamefont {P.~N.}\ \bibnamefont
  {Hai}}, \bibinfo {author} {\bibfnamefont {L.~D.}\ \bibnamefont {Anh}},
  \bibinfo {author} {\bibfnamefont {S.}~\bibnamefont {Mohan}}, \bibinfo
  {author} {\bibfnamefont {T.}~\bibnamefont {Tamegai}}, \bibinfo {author}
  {\bibfnamefont {M.}~\bibnamefont {Kodzuka}}, \bibinfo {author} {\bibfnamefont
  {T.}~\bibnamefont {Ohkubo}}, \bibinfo {author} {\bibfnamefont
  {K.}~\bibnamefont {Hono}}, \ and\ \bibinfo {author} {\bibfnamefont
  {M.}~\bibnamefont {Tanaka}},\ }\href {\doibase
  http://dx.doi.org/10.1063/1.4764947} {\bibfield  {journal} {\bibinfo
  {journal} {Applied Physics Letters}\ }\textbf {\bibinfo {volume} {101}},\
  \bibinfo {eid} {182403} (\bibinfo {year} {2012}{\natexlab{b}})}\BibitemShut
  {NoStop}%
\bibitem [{\citenamefont {Hai}\ \emph {et~al.}(2012{\natexlab{c}})\citenamefont
  {Hai}, \citenamefont {Anh},\ and\ \citenamefont {Tanaka}}]{Nam-Hai:2012ab}%
  \BibitemOpen
  \bibfield  {author} {\bibinfo {author} {\bibfnamefont {P.~N.}\ \bibnamefont
  {Hai}}, \bibinfo {author} {\bibfnamefont {L.~D.}\ \bibnamefont {Anh}}, \ and\
  \bibinfo {author} {\bibfnamefont {M.}~\bibnamefont {Tanaka}},\ }\href
  {\doibase http://dx.doi.org/10.1063/1.4772630} {\bibfield  {journal}
  {\bibinfo  {journal} {Appl. Phys. Lett.}\ }\textbf {\bibinfo {volume}
  {101}},\ \bibinfo {eid} {252410} (\bibinfo {year}
  {2012}{\natexlab{c}})}\BibitemShut {NoStop}%
\bibitem [{\citenamefont {Anh}\ \emph {et~al.}(2014)\citenamefont {Anh},
  \citenamefont {Hai},\ and\ \citenamefont {Tanaka}}]{Duc-Anh:2014aa}%
  \BibitemOpen
  \bibfield  {author} {\bibinfo {author} {\bibfnamefont {L.~D.}\ \bibnamefont
  {Anh}}, \bibinfo {author} {\bibfnamefont {P.~N.}\ \bibnamefont {Hai}}, \ and\
  \bibinfo {author} {\bibfnamefont {M.}~\bibnamefont {Tanaka}},\ }\href
  {\doibase http://dx.doi.org/10.1063/1.4863214} {\bibfield  {journal}
  {\bibinfo  {journal} {Appl. Phys. Lett.}\ }\textbf {\bibinfo {volume}
  {104}},\ \bibinfo {eid} {042404} (\bibinfo {year} {2014})}\BibitemShut
  {NoStop}%
\bibitem [{\citenamefont {Sasaki}\ \emph {et~al.}(2014)\citenamefont {Sasaki},
  \citenamefont {Anh}, \citenamefont {Hai},\ and\ \citenamefont
  {Tanaka}}]{Sasaki:2014aa}%
  \BibitemOpen
  \bibfield  {author} {\bibinfo {author} {\bibfnamefont {D.}~\bibnamefont
  {Sasaki}}, \bibinfo {author} {\bibfnamefont {L.~D.}\ \bibnamefont {Anh}},
  \bibinfo {author} {\bibfnamefont {P.~N.}\ \bibnamefont {Hai}}, \ and\
  \bibinfo {author} {\bibfnamefont {M.}~\bibnamefont {Tanaka}},\ }\href
  {\doibase http://dx.doi.org/10.1063/1.4870970} {\bibfield  {journal}
  {\bibinfo  {journal} {Appl. Phys. Lett.}\ }\textbf {\bibinfo {volume}
  {104}},\ \bibinfo {eid} {142406} (\bibinfo {year} {2014})}\BibitemShut
  {NoStop}%
\bibitem [{\citenamefont {Haneda}\ \emph {et~al.}(2000)\citenamefont {Haneda},
  \citenamefont {Yamaura}, \citenamefont {Takatani}, \citenamefont {Hara},
  \citenamefont {ichi Harigae},\ and\ \citenamefont
  {Munekata}}]{Haneda:2000aa}%
  \BibitemOpen
  \bibfield  {author} {\bibinfo {author} {\bibfnamefont {S.}~\bibnamefont
  {Haneda}}, \bibinfo {author} {\bibfnamefont {M.}~\bibnamefont {Yamaura}},
  \bibinfo {author} {\bibfnamefont {Y.}~\bibnamefont {Takatani}}, \bibinfo
  {author} {\bibfnamefont {K.}~\bibnamefont {Hara}}, \bibinfo {author}
  {\bibfnamefont {S.}~\bibnamefont {ichi Harigae}}, \ and\ \bibinfo {author}
  {\bibfnamefont {H.}~\bibnamefont {Munekata}},\ }\href
  {http://stacks.iop.org/1347-4065/39/i=1A/a=L9} {\bibfield  {journal}
  {\bibinfo  {journal} {Jpn. J. Appl. Phys.}\ }\textbf {\bibinfo {volume}
  {39}},\ \bibinfo {pages} {L9} (\bibinfo {year} {2000})}\BibitemShut {NoStop}%
\bibitem [{\citenamefont {Kobayashi}\ \emph {et~al.}(2010)\citenamefont
  {Kobayashi}, \citenamefont {Ishida}, \citenamefont {Hwang}, \citenamefont
  {Osafune}, \citenamefont {Fujimori}, \citenamefont {Takeda}, \citenamefont
  {Okane}, \citenamefont {Saitoh}, \citenamefont {Kobayashi}, \citenamefont
  {Saeki}, \citenamefont {Kawai},\ and\ \citenamefont
  {Tabata}}]{kobayashi:2010aa}%
  \BibitemOpen
  \bibfield  {author} {\bibinfo {author} {\bibfnamefont {M.}~\bibnamefont
  {Kobayashi}}, \bibinfo {author} {\bibfnamefont {Y.}~\bibnamefont {Ishida}},
  \bibinfo {author} {\bibfnamefont {J.~I.}\ \bibnamefont {Hwang}}, \bibinfo
  {author} {\bibfnamefont {Y.}~\bibnamefont {Osafune}}, \bibinfo {author}
  {\bibfnamefont {A.}~\bibnamefont {Fujimori}}, \bibinfo {author}
  {\bibfnamefont {Y.}~\bibnamefont {Takeda}}, \bibinfo {author} {\bibfnamefont
  {T.}~\bibnamefont {Okane}}, \bibinfo {author} {\bibfnamefont
  {Y.}~\bibnamefont {Saitoh}}, \bibinfo {author} {\bibfnamefont
  {K.}~\bibnamefont {Kobayashi}}, \bibinfo {author} {\bibfnamefont
  {H.}~\bibnamefont {Saeki}}, \bibinfo {author} {\bibfnamefont
  {T.}~\bibnamefont {Kawai}}, \ and\ \bibinfo {author} {\bibfnamefont
  {H.}~\bibnamefont {Tabata}},\ }\href {\doibase 10.1103/PhysRevB.81.075204}
  {\bibfield  {journal} {\bibinfo  {journal} {Phys. Rev. B}\ }\textbf {\bibinfo
  {volume} {81}},\ \bibinfo {pages} {075204} (\bibinfo {year}
  {2010})}\BibitemShut {NoStop}%
\bibitem [{\citenamefont {Carra}\ \emph {et~al.}(1993)\citenamefont {Carra},
  \citenamefont {Thole}, \citenamefont {Altarelli},\ and\ \citenamefont
  {Wang}}]{carra:1993aa}%
  \BibitemOpen
  \bibfield  {author} {\bibinfo {author} {\bibfnamefont {P.}~\bibnamefont
  {Carra}}, \bibinfo {author} {\bibfnamefont {B.~T.}\ \bibnamefont {Thole}},
  \bibinfo {author} {\bibfnamefont {M.}~\bibnamefont {Altarelli}}, \ and\
  \bibinfo {author} {\bibfnamefont {X.}~\bibnamefont {Wang}},\ }\href {\doibase
  10.1103/PhysRevLett.70.694} {\bibfield  {journal} {\bibinfo  {journal} {Phys.
  Rev. Lett.}\ }\textbf {\bibinfo {volume} {70}},\ \bibinfo {pages} {694}
  (\bibinfo {year} {1993})}\BibitemShut {NoStop}%
\bibitem [{\citenamefont {Thole}\ \emph {et~al.}(1992)\citenamefont {Thole},
  \citenamefont {Carra}, \citenamefont {Sette},\ and\ \citenamefont {van~der
  Laan}}]{thole:1992aa}%
  \BibitemOpen
  \bibfield  {author} {\bibinfo {author} {\bibfnamefont {B.~T.}\ \bibnamefont
  {Thole}}, \bibinfo {author} {\bibfnamefont {P.}~\bibnamefont {Carra}},
  \bibinfo {author} {\bibfnamefont {F.}~\bibnamefont {Sette}}, \ and\ \bibinfo
  {author} {\bibfnamefont {G.}~\bibnamefont {van~der Laan}},\ }\href {\doibase
  10.1103/PhysRevLett.68.1943} {\bibfield  {journal} {\bibinfo  {journal}
  {Phys. Rev. Lett.}\ }\textbf {\bibinfo {volume} {68}},\ \bibinfo {pages}
  {1943} (\bibinfo {year} {1992})}\BibitemShut {NoStop}%
\bibitem [{\citenamefont {Kobayashi}\ \emph {et~al.}(2014)\citenamefont
  {Kobayashi}, \citenamefont {Anh}, \citenamefont {Hai}, \citenamefont
  {Takeda}, \citenamefont {Sakamoto}, \citenamefont {Kadono}, \citenamefont
  {Okane}, \citenamefont {Saitoh}, \citenamefont {Yamagami}, \citenamefont
  {Harada}, \citenamefont {Oshima}, \citenamefont {Tanaka},\ and\ \citenamefont
  {Fujimori}}]{kobayashi:2014aa}%
  \BibitemOpen
  \bibfield  {author} {\bibinfo {author} {\bibfnamefont {M.}~\bibnamefont
  {Kobayashi}}, \bibinfo {author} {\bibfnamefont {L.~D.}\ \bibnamefont {Anh}},
  \bibinfo {author} {\bibfnamefont {P.~N.}\ \bibnamefont {Hai}}, \bibinfo
  {author} {\bibfnamefont {Y.}~\bibnamefont {Takeda}}, \bibinfo {author}
  {\bibfnamefont {S.}~\bibnamefont {Sakamoto}}, \bibinfo {author}
  {\bibfnamefont {T.}~\bibnamefont {Kadono}}, \bibinfo {author} {\bibfnamefont
  {T.}~\bibnamefont {Okane}}, \bibinfo {author} {\bibfnamefont
  {Y.}~\bibnamefont {Saitoh}}, \bibinfo {author} {\bibfnamefont
  {H.}~\bibnamefont {Yamagami}}, \bibinfo {author} {\bibfnamefont
  {Y.}~\bibnamefont {Harada}}, \bibinfo {author} {\bibfnamefont
  {M.}~\bibnamefont {Oshima}}, \bibinfo {author} {\bibfnamefont
  {M.}~\bibnamefont {Tanaka}}, \ and\ \bibinfo {author} {\bibfnamefont
  {A.}~\bibnamefont {Fujimori}},\ }\href {\doibase
  http://dx.doi.org/10.1063/1.4890733} {\bibfield  {journal} {\bibinfo
  {journal} {Appl. Phys. Lett.}\ }\textbf {\bibinfo {volume} {105}},\ \bibinfo
  {eid} {032403} (\bibinfo {year} {2014})}\BibitemShut {NoStop}%
\bibitem [{\citenamefont {Chen}\ \emph {et~al.}(1995)\citenamefont {Chen},
  \citenamefont {Idzerda}, \citenamefont {Lin}, \citenamefont {Smith},
  \citenamefont {Meigs}, \citenamefont {Chaban}, \citenamefont {Ho},
  \citenamefont {Pellegrin},\ and\ \citenamefont {Sette}}]{Chen:1995aa}%
  \BibitemOpen
  \bibfield  {author} {\bibinfo {author} {\bibfnamefont {C.~T.}\ \bibnamefont
  {Chen}}, \bibinfo {author} {\bibfnamefont {Y.~U.}\ \bibnamefont {Idzerda}},
  \bibinfo {author} {\bibfnamefont {H.-J.}\ \bibnamefont {Lin}}, \bibinfo
  {author} {\bibfnamefont {N.~V.}\ \bibnamefont {Smith}}, \bibinfo {author}
  {\bibfnamefont {G.}~\bibnamefont {Meigs}}, \bibinfo {author} {\bibfnamefont
  {E.}~\bibnamefont {Chaban}}, \bibinfo {author} {\bibfnamefont {G.~H.}\
  \bibnamefont {Ho}}, \bibinfo {author} {\bibfnamefont {E.}~\bibnamefont
  {Pellegrin}}, \ and\ \bibinfo {author} {\bibfnamefont {F.}~\bibnamefont
  {Sette}},\ }\href {\doibase 10.1103/PhysRevLett.75.152} {\bibfield  {journal}
  {\bibinfo  {journal} {Phys. Rev. Lett.}\ }\textbf {\bibinfo {volume} {75}},\
  \bibinfo {pages} {152} (\bibinfo {year} {1995})}\BibitemShut {NoStop}%
\bibitem [{\citenamefont {Brice-Profeta}\ \emph {et~al.}(2005)\citenamefont
  {Brice-Profeta}, \citenamefont {Arrio}, \citenamefont {Tronc}, \citenamefont
  {Menguy}, \citenamefont {Letard}, \citenamefont {dit Moulin}, \citenamefont
  {Nogu{\`e}s}, \citenamefont {Chan{\'e}ac}, \citenamefont {Jolivet},\ and\
  \citenamefont {Sainctavit}}]{BriceProfeta:2005aa}%
  \BibitemOpen
  \bibfield  {author} {\bibinfo {author} {\bibfnamefont {S.}~\bibnamefont
  {Brice-Profeta}}, \bibinfo {author} {\bibfnamefont {M.-A.}\ \bibnamefont
  {Arrio}}, \bibinfo {author} {\bibfnamefont {E.}~\bibnamefont {Tronc}},
  \bibinfo {author} {\bibfnamefont {N.}~\bibnamefont {Menguy}}, \bibinfo
  {author} {\bibfnamefont {I.}~\bibnamefont {Letard}}, \bibinfo {author}
  {\bibfnamefont {C.~C.}\ \bibnamefont {dit Moulin}}, \bibinfo {author}
  {\bibfnamefont {M.}~\bibnamefont {Nogu{\`e}s}}, \bibinfo {author}
  {\bibfnamefont {C.}~\bibnamefont {Chan{\'e}ac}}, \bibinfo {author}
  {\bibfnamefont {J.-P.}\ \bibnamefont {Jolivet}}, \ and\ \bibinfo {author}
  {\bibfnamefont {P.}~\bibnamefont {Sainctavit}},\ }\href {\doibase
  http://dx.doi.org/10.1016/j.jmmm.2004.09.120} {\bibfield  {journal} {\bibinfo
   {journal} {J. Magn. Magn. Mater.}\ }\textbf {\bibinfo {volume} {288}},\
  \bibinfo {pages} {354 } (\bibinfo {year} {2005})}\BibitemShut {NoStop}%
\bibitem [{\citenamefont {Takeda}\ \emph {et~al.}(2008)\citenamefont {Takeda},
  \citenamefont {Kobayashi}, \citenamefont {Okane}, \citenamefont {Ohkochi},
  \citenamefont {Okamoto}, \citenamefont {Saitoh}, \citenamefont {Kobayashi},
  \citenamefont {Yamagami}, \citenamefont {Fujimori}, \citenamefont {Tanaka},
  \citenamefont {Okabayashi}, \citenamefont {Oshima}, \citenamefont {Ohya},
  \citenamefont {Hai},\ and\ \citenamefont {Tanaka}}]{takeda:2008aa}%
  \BibitemOpen
  \bibfield  {author} {\bibinfo {author} {\bibfnamefont {Y.}~\bibnamefont
  {Takeda}}, \bibinfo {author} {\bibfnamefont {M.}~\bibnamefont {Kobayashi}},
  \bibinfo {author} {\bibfnamefont {T.}~\bibnamefont {Okane}}, \bibinfo
  {author} {\bibfnamefont {T.}~\bibnamefont {Ohkochi}}, \bibinfo {author}
  {\bibfnamefont {J.}~\bibnamefont {Okamoto}}, \bibinfo {author} {\bibfnamefont
  {Y.}~\bibnamefont {Saitoh}}, \bibinfo {author} {\bibfnamefont
  {K.}~\bibnamefont {Kobayashi}}, \bibinfo {author} {\bibfnamefont
  {H.}~\bibnamefont {Yamagami}}, \bibinfo {author} {\bibfnamefont
  {A.}~\bibnamefont {Fujimori}}, \bibinfo {author} {\bibfnamefont
  {A.}~\bibnamefont {Tanaka}}, \bibinfo {author} {\bibfnamefont
  {J.}~\bibnamefont {Okabayashi}}, \bibinfo {author} {\bibfnamefont
  {M.}~\bibnamefont {Oshima}}, \bibinfo {author} {\bibfnamefont
  {S.}~\bibnamefont {Ohya}}, \bibinfo {author} {\bibfnamefont {P.~N.}\
  \bibnamefont {Hai}}, \ and\ \bibinfo {author} {\bibfnamefont
  {M.}~\bibnamefont {Tanaka}},\ }\href {\doibase
  10.1103/PhysRevLett.100.247202} {\bibfield  {journal} {\bibinfo  {journal}
  {Phys. Rev. Lett.}\ }\textbf {\bibinfo {volume} {100}},\ \bibinfo {pages}
  {247202} (\bibinfo {year} {2008})}\BibitemShut {NoStop}%
\bibitem [{\citenamefont {van~der Laan}\ and\ \citenamefont
  {Kirkman}(1992)}]{Laan:1992aa}%
  \BibitemOpen
  \bibfield  {author} {\bibinfo {author} {\bibfnamefont {G.}~\bibnamefont
  {van~der Laan}}\ and\ \bibinfo {author} {\bibfnamefont {I.~W.}\ \bibnamefont
  {Kirkman}},\ }\href {http://stacks.iop.org/0953-8984/4/i=16/a=019} {\bibfield
   {journal} {\bibinfo  {journal} {J. Phys.: Condens. Matter}\ }\textbf
  {\bibinfo {volume} {4}},\ \bibinfo {pages} {4189} (\bibinfo {year}
  {1992})}\BibitemShut {NoStop}%
\bibitem [{\citenamefont {Kang}\ \emph {et~al.}(2008)\citenamefont {Kang},
  \citenamefont {Kim}, \citenamefont {Lee}, \citenamefont {Kim}, \citenamefont
  {Kim}, \citenamefont {Han}, \citenamefont {Kim}, \citenamefont {Kim},
  \citenamefont {Lee}, \citenamefont {Kim},\ and\ \citenamefont
  {Min}}]{kang:2008aa}%
  \BibitemOpen
  \bibfield  {author} {\bibinfo {author} {\bibfnamefont {J.-S.}\ \bibnamefont
  {Kang}}, \bibinfo {author} {\bibfnamefont {G.}~\bibnamefont {Kim}}, \bibinfo
  {author} {\bibfnamefont {H.~J.}\ \bibnamefont {Lee}}, \bibinfo {author}
  {\bibfnamefont {H.~S.}\ \bibnamefont {Kim}}, \bibinfo {author} {\bibfnamefont
  {D.~H.}\ \bibnamefont {Kim}}, \bibinfo {author} {\bibfnamefont {S.~W.}\
  \bibnamefont {Han}}, \bibinfo {author} {\bibfnamefont {S.~J.}\ \bibnamefont
  {Kim}}, \bibinfo {author} {\bibfnamefont {C.~S.}\ \bibnamefont {Kim}},
  \bibinfo {author} {\bibfnamefont {H.}~\bibnamefont {Lee}}, \bibinfo {author}
  {\bibfnamefont {J.-Y.}\ \bibnamefont {Kim}}, \ and\ \bibinfo {author}
  {\bibfnamefont {B.~I.}\ \bibnamefont {Min}},\ }\href {\doibase
  http://dx.doi.org/10.1063/1.2839617} {\bibfield  {journal} {\bibinfo
  {journal} {J. Appl. Phys.}\ }\textbf {\bibinfo {volume} {103}},\  (\bibinfo
  {year} {2008})}\BibitemShut {NoStop}%
\bibitem [{\citenamefont {Kowalik}\ \emph {et~al.}(2012)\citenamefont
  {Kowalik}, \citenamefont {Persson}, \citenamefont {Ni\~no}, \citenamefont
  {Navarro-Quezada}, \citenamefont {Faina}, \citenamefont {Bonanni},
  \citenamefont {Dietl},\ and\ \citenamefont {Arvanitis}}]{kowalik:2012aa}%
  \BibitemOpen
  \bibfield  {author} {\bibinfo {author} {\bibfnamefont {I.~A.}\ \bibnamefont
  {Kowalik}}, \bibinfo {author} {\bibfnamefont {A.}~\bibnamefont {Persson}},
  \bibinfo {author} {\bibfnamefont {M.~A.}\ \bibnamefont {Ni\~no}}, \bibinfo
  {author} {\bibfnamefont {A.}~\bibnamefont {Navarro-Quezada}}, \bibinfo
  {author} {\bibfnamefont {B.}~\bibnamefont {Faina}}, \bibinfo {author}
  {\bibfnamefont {A.}~\bibnamefont {Bonanni}}, \bibinfo {author} {\bibfnamefont
  {T.}~\bibnamefont {Dietl}}, \ and\ \bibinfo {author} {\bibfnamefont
  {D.}~\bibnamefont {Arvanitis}},\ }\href {\doibase 10.1103/PhysRevB.85.184411}
  {\bibfield  {journal} {\bibinfo  {journal} {Phys. Rev. B}\ }\textbf {\bibinfo
  {volume} {85}},\ \bibinfo {pages} {184411} (\bibinfo {year}
  {2012})}\BibitemShut {NoStop}%
\bibitem [{\citenamefont {Saitoh}\ \emph {et~al.}(2012)\citenamefont {Saitoh},
  \citenamefont {Fukuda}, \citenamefont {Takeda}, \citenamefont {Yamagami},
  \citenamefont {Takahashi}, \citenamefont {Asano}, \citenamefont {Hara},
  \citenamefont {Shirasawa}, \citenamefont {Takeuchi}, \citenamefont {Tanaka},\
  and\ \citenamefont {Hideo}}]{Saitoh:2012aa}%
  \BibitemOpen
  \bibfield  {author} {\bibinfo {author} {\bibfnamefont {Y.}~\bibnamefont
  {Saitoh}}, \bibinfo {author} {\bibfnamefont {Y.}~\bibnamefont {Fukuda}},
  \bibinfo {author} {\bibfnamefont {Y.}~\bibnamefont {Takeda}}, \bibinfo
  {author} {\bibfnamefont {H.}~\bibnamefont {Yamagami}}, \bibinfo {author}
  {\bibfnamefont {S.}~\bibnamefont {Takahashi}}, \bibinfo {author}
  {\bibfnamefont {Y.}~\bibnamefont {Asano}}, \bibinfo {author} {\bibfnamefont
  {T.}~\bibnamefont {Hara}}, \bibinfo {author} {\bibfnamefont {K.}~\bibnamefont
  {Shirasawa}}, \bibinfo {author} {\bibfnamefont {M.}~\bibnamefont {Takeuchi}},
  \bibinfo {author} {\bibfnamefont {T.}~\bibnamefont {Tanaka}}, \ and\ \bibinfo
  {author} {\bibfnamefont {K.}~\bibnamefont {Hideo}},\ }\href@noop {}
  {\bibfield  {journal} {\bibinfo  {journal} {Journal of synchrotron
  radiation}\ }\textbf {\bibinfo {volume} {19}},\ \bibinfo {pages} {388}
  (\bibinfo {year} {2012})}\BibitemShut {NoStop}%
\bibitem [{\citenamefont {St\"ohr}\ and\ \citenamefont
  {K\"onig}(1995)}]{stohr:1995aa}%
  \BibitemOpen
  \bibfield  {author} {\bibinfo {author} {\bibfnamefont {J.}~\bibnamefont
  {St\"ohr}}\ and\ \bibinfo {author} {\bibfnamefont {H.}~\bibnamefont
  {K\"onig}},\ }\href {\doibase 10.1103/PhysRevLett.75.3748} {\bibfield
  {journal} {\bibinfo  {journal} {Phys. Rev. Lett.}\ }\textbf {\bibinfo
  {volume} {75}},\ \bibinfo {pages} {3748} (\bibinfo {year}
  {1995})}\BibitemShut {NoStop}%
\bibitem [{\citenamefont {Piamonteze}\ \emph {et~al.}(2009)\citenamefont
  {Piamonteze}, \citenamefont {Miedema},\ and\ \citenamefont
  {de~Groot}}]{Piamonteze:2009aa}%
  \BibitemOpen
  \bibfield  {author} {\bibinfo {author} {\bibfnamefont {C.}~\bibnamefont
  {Piamonteze}}, \bibinfo {author} {\bibfnamefont {P.}~\bibnamefont {Miedema}},
  \ and\ \bibinfo {author} {\bibfnamefont {F.~M.~F.}\ \bibnamefont
  {de~Groot}},\ }\href {\doibase 10.1103/PhysRevB.80.184410} {\bibfield
  {journal} {\bibinfo  {journal} {Phys. Rev. B}\ }\textbf {\bibinfo {volume}
  {80}},\ \bibinfo {pages} {184410} (\bibinfo {year} {2009})}\BibitemShut
  {NoStop}%
\bibitem [{\citenamefont {Kaminski}\ and\ \citenamefont
  {Das~Sarma}(2002)}]{kaminski:2002aa}%
  \BibitemOpen
  \bibfield  {author} {\bibinfo {author} {\bibfnamefont {A.}~\bibnamefont
  {Kaminski}}\ and\ \bibinfo {author} {\bibfnamefont {S.}~\bibnamefont
  {Das~Sarma}},\ }\href {\doibase 10.1103/PhysRevLett.88.247202} {\bibfield
  {journal} {\bibinfo  {journal} {Phys. Rev. Lett.}\ }\textbf {\bibinfo
  {volume} {88}},\ \bibinfo {pages} {247202} (\bibinfo {year}
  {2002})}\BibitemShut {NoStop}%
\bibitem [{\citenamefont {Coey}\ \emph {et~al.}(2005)\citenamefont {Coey},
  \citenamefont {Venkatesan},\ and\ \citenamefont {Fitzgerald}}]{coey:2005aa}%
  \BibitemOpen
  \bibfield  {author} {\bibinfo {author} {\bibfnamefont {J.~M.~D.}\
  \bibnamefont {Coey}}, \bibinfo {author} {\bibfnamefont {M.}~\bibnamefont
  {Venkatesan}}, \ and\ \bibinfo {author} {\bibfnamefont {C.~B.}\ \bibnamefont
  {Fitzgerald}},\ }\href {http://dx.doi.org/10.1038/nmat1310} {\bibfield
  {journal} {\bibinfo  {journal} {Nat. Mater.}\ }\textbf {\bibinfo {volume}
  {4}},\ \bibinfo {pages} {173} (\bibinfo {year} {2005})}\BibitemShut {NoStop}%
\bibitem [{\citenamefont {Sato}\ \emph {et~al.}(2005)\citenamefont {Sato},
  \citenamefont {Katayama-Yoshida},\ and\ \citenamefont
  {Dederichs}}]{sato:2005aa}%
  \BibitemOpen
  \bibfield  {author} {\bibinfo {author} {\bibfnamefont {K.}~\bibnamefont
  {Sato}}, \bibinfo {author} {\bibfnamefont {H.}~\bibnamefont
  {Katayama-Yoshida}}, \ and\ \bibinfo {author} {\bibfnamefont {P.~H.}\
  \bibnamefont {Dederichs}},\ }\href
  {http://stacks.iop.org/1347-4065/44/i=7L/a=L948} {\bibfield  {journal}
  {\bibinfo  {journal} {Jpn. J. Appl. Phys.}\ }\textbf {\bibinfo {volume}
  {44}},\ \bibinfo {pages} {L948} (\bibinfo {year} {2005})}\BibitemShut
  {NoStop}%
\bibitem [{\citenamefont {Kuroda}\ \emph {et~al.}(2007)\citenamefont {Kuroda},
  \citenamefont {Nishizawa}, \citenamefont {Takita}, \citenamefont {Mitome},
  \citenamefont {Bando}, \citenamefont {Osuch},\ and\ \citenamefont
  {Dietl}}]{kuroda:2007aa}%
  \BibitemOpen
  \bibfield  {author} {\bibinfo {author} {\bibfnamefont {S.}~\bibnamefont
  {Kuroda}}, \bibinfo {author} {\bibfnamefont {N.}~\bibnamefont {Nishizawa}},
  \bibinfo {author} {\bibfnamefont {K.}~\bibnamefont {Takita}}, \bibinfo
  {author} {\bibfnamefont {M.}~\bibnamefont {Mitome}}, \bibinfo {author}
  {\bibfnamefont {Y.}~\bibnamefont {Bando}}, \bibinfo {author} {\bibfnamefont
  {K.}~\bibnamefont {Osuch}}, \ and\ \bibinfo {author} {\bibfnamefont
  {T.}~\bibnamefont {Dietl}},\ }\href {http://dx.doi.org/10.1038/nmat1910}
  {\bibfield  {journal} {\bibinfo  {journal} {Nat. Mater.}\ }\textbf {\bibinfo
  {volume} {6}},\ \bibinfo {pages} {440} (\bibinfo {year} {2007})}\BibitemShut
  {NoStop}%
\bibitem [{\citenamefont {Dietl}\ \emph {et~al.}(2014)\citenamefont {Dietl},
  \citenamefont {Sato}, \citenamefont {Fukushima}, \citenamefont {Bonanni},
  \citenamefont {Jamet}, \citenamefont {Barski}, \citenamefont {Kuroda},
  \citenamefont {Tanaka}, \citenamefont {Hai},\ and\ \citenamefont
  {Katayama-Yoshida}}]{dietl:2014ab}%
  \BibitemOpen
  \bibfield  {author} {\bibinfo {author} {\bibfnamefont {T.}~\bibnamefont
  {Dietl}}, \bibinfo {author} {\bibfnamefont {K.}~\bibnamefont {Sato}},
  \bibinfo {author} {\bibfnamefont {T.}~\bibnamefont {Fukushima}}, \bibinfo
  {author} {\bibfnamefont {A.}~\bibnamefont {Bonanni}}, \bibinfo {author}
  {\bibfnamefont {M.}~\bibnamefont {Jamet}}, \bibinfo {author} {\bibfnamefont
  {A.}~\bibnamefont {Barski}}, \bibinfo {author} {\bibfnamefont
  {S.}~\bibnamefont {Kuroda}}, \bibinfo {author} {\bibfnamefont
  {M.}~\bibnamefont {Tanaka}}, \bibinfo {author} {\bibfnamefont {P.~N.}\
  \bibnamefont {Hai}}, \ and\ \bibinfo {author} {\bibfnamefont
  {H.}~\bibnamefont {Katayama-Yoshida}},\ }\href@noop {} {\bibfield  {journal}
  {\bibinfo  {journal} {ArXiv:1412.8062v2}\ } (\bibinfo {year}
  {2014})}\BibitemShut {NoStop}%
\bibitem [{\citenamefont {Coey}\ \emph {et~al.}(2010)\citenamefont {Coey},
  \citenamefont {Mlack}, \citenamefont {Venkatesan},\ and\ \citenamefont
  {Stamenov}}]{coey:2010aa}%
  \BibitemOpen
  \bibfield  {author} {\bibinfo {author} {\bibfnamefont {J.~M.~D.}\
  \bibnamefont {Coey}}, \bibinfo {author} {\bibfnamefont {J.}~\bibnamefont
  {Mlack}}, \bibinfo {author} {\bibfnamefont {M.}~\bibnamefont {Venkatesan}}, \
  and\ \bibinfo {author} {\bibfnamefont {P.}~\bibnamefont {Stamenov}},\ }\href
  {\doibase 10.1109/TMAG.2010.2041910} {\bibfield  {journal} {\bibinfo
  {journal} {IEEE Trans. Magn.}\ }\textbf {\bibinfo {volume} {46}},\ \bibinfo
  {pages} {2501} (\bibinfo {year} {2010})}\BibitemShut {NoStop}%
\end{thebibliography}%

\end{document}